\documentclass{article}

\usepackage{authblk}
\usepackage{amssymb}
\usepackage[backend=biber,style=chem-acs]{biblatex}
\usepackage{booktabs}
\usepackage[margin=1in]{geometry}
\usepackage{graphicx}
\usepackage{multirow}
\usepackage{placeins}
\usepackage{siunitx}
\usepackage{threeparttable}
\usepackage{url}
\usepackage{xcolor}

\usepackage{hyperref}
\usepackage{cleveref}
\newcommand*\therepo{https://github.com/coleygroup/rogi-xd}
\newcommand*\thetitle{Evaluating the roughness of structure-property relationships using pretrained molecular representations}

\newcommand*\ccb{Department of Chemistry and Chemical Biology, Harvard University, Cambridge, MA 02138, United States}
\newcommand*\cheme{Department of Chemical Engineering, MIT, Cambridge, MA 02139, United States}
\newcommand*\ibm{IBM Research Europe, Warrington WA4 4AD, United Kingdom}
\newcommand*\ibmtwo{IBM Thomas J. Watson Research Center, Cambridge, MA 02142, United States}

\newcommand*\eecs{Department of Electrical Engineering and Computer Science, MIT, Cambridge, MA 02139, United States}

\newcommand*\rogiOrig{\textbf{RO}u\textbf{G}hness \textbf{I}ndex}
\newcommand*\newrogi{ROGI-XD}

\newcommand*\vae{variational autoencoder}
\newcommand*\gin{graph isomorphism network}
\newcommand*\fp{Morgan fingerprint}

\newcommand*\knn{$k$-nearest neighbors}
\newcommand*\mlp{multilayer perceptron}
\newcommand*\pls{partial least squares}
\newcommand*\rf{random forest}
\newcommand*\svr{support vector regression}

\newcommand*\reprCaption{\textit{FP}: \fp; \textit{VAE}: \vae; \textit{GIN}: \gin}
\newcommand*\modelCaption{
    \textit{KNN}: \knn; \textit{MLP}: \mlp; \textit{PLS}: \pls; \textit{RF}: \rf; \textit{SVR}: \svr
}

\newcommand*{\range}[2]{\left[#1,\,#2\right]}

\newcommand{\tcite}[1]{\citeauthor{#1}\autocite{#1}}

\makeatletter
\newcommand*\settitle\@maketitle
\makeatother

\title{\thetitle}

\author[1,2]{David E. Graff}
\author[3]{Edward O. Pyzer-Knapp}
\author[4]{Kirk E. Jordan}
\author[1]{Eugene I. Shakhnovich}
\author[2,5]{Connor W. Coley}

\affil[1]{\ccb}
\affil[2]{\cheme}
\affil[3]{\ibm}
\affil[4]{\ibmtwo}
\affil[5]{\eecs}

\date{}

\begin{document}

\maketitle

\section*{Abstract}

Quantitative structure-property relationships (QSPRs) aid in understanding molecular properties as a function of molecular structure. When the correlation between structure and property weakens, a dataset is described as ``rough,'' but this characteristic is partly a function of the chosen representation. Among possible molecular representations are those from recently-developed ``foundation models'' for chemistry which learn molecular representation from unlabeled samples via self-supervision. However, the performance of these pretrained representations on property prediction benchmarks is mixed when compared to baseline approaches. We sought to understand these trends in terms of the roughness of the underlying QSPR surfaces. We introduce a reformulation of the roughness index (ROGI), \newrogi{}, to enable comparison of ROGI values across representations and evaluate various pretrained representations and those constructed by simple fingerprints and descriptors. We show that pretrained representations do not produce smoother QSPR surfaces, in agreement with previous empirical results of model accuracy. Our findings suggest that imposing stronger assumptions of smoothness with respect to molecular structure during model pretraining can aid in the downstream generation of smoother QSPR surfaces.

\section*{Introduction}

\begin{figure}[t!]
    \centering
    \includegraphics[width=0.9\textwidth]{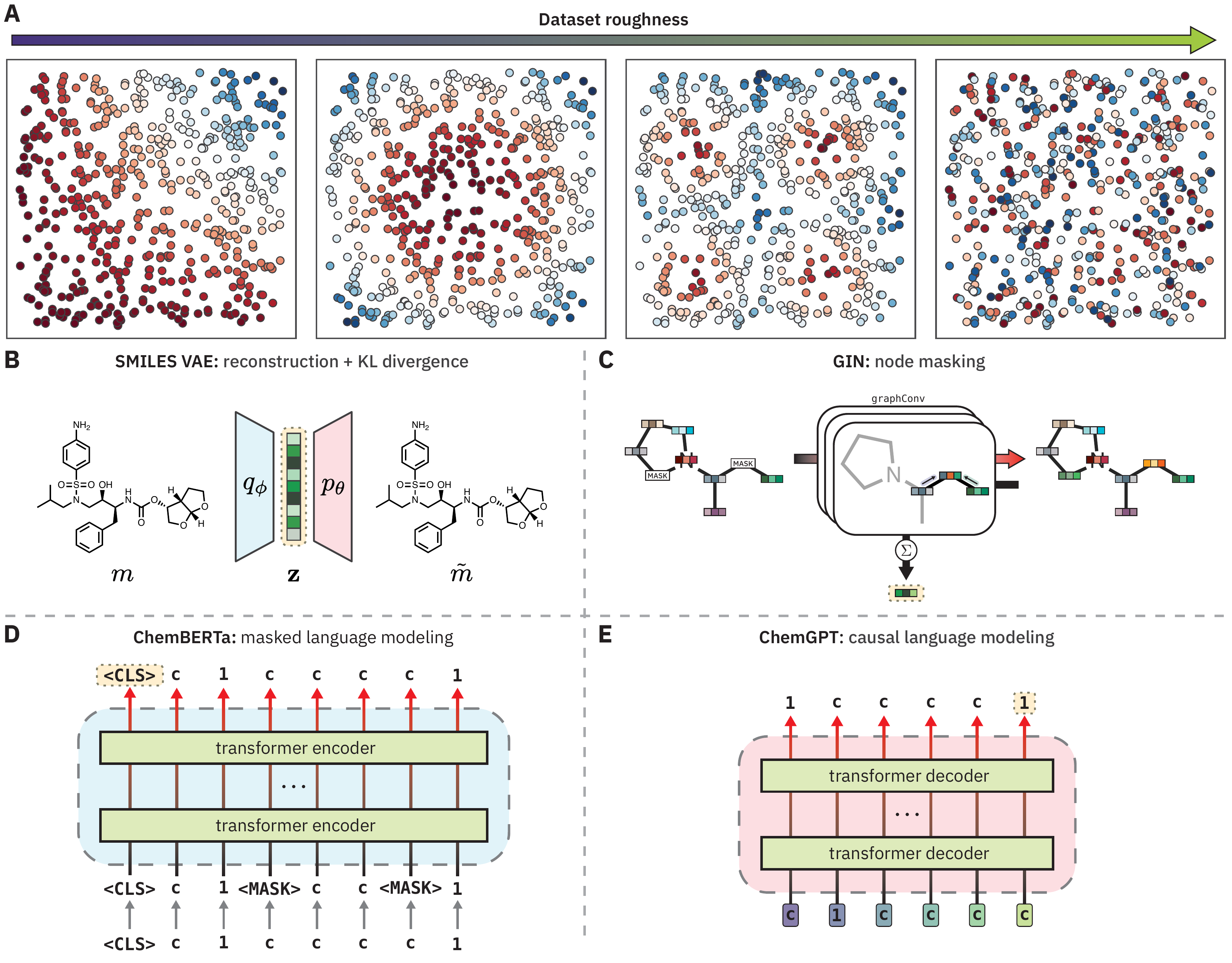}
    \caption{
        \textbf{(A)} Examples of increasingly rough datasets.
        \textbf{(B, C, D, E)} Schematics of pretrained chemical models evaluated in this study. Yellow, dotted boxes indicate the source of the pretrained representation.
        \textbf{(B)} A SMILES \vae{} (VAE) uses the mean latent representation after encoding.
        \textbf{(C)} A \gin{} (GIN) uses the sum of the node hidden representations after graph convolutions.
        \textbf{(D)} ChemBERTa uses the embedding of the \texttt{<CLS>} token after the final transformer encoder block.
        \textbf{(E)} ChemGPT uses the embedding of the \emph{last} token in the input sequence after the final transformer decoder block.
    }
    \label{fig:schematic}
\end{figure}

The development of quantitative structure-property relationships (QSPRs) is central to molecular discovery, as they help rationalize trends in molecular properties and suggest to chemists how they can or should modify certain structural motifs to achieve a target property. A key challenge in building QSPRs arises when similar molecules possess divergent property labels. These scenarios, so-called ``activity cliffs,''\autocite{silipo_qsar_1991,maggiora_outliers_2006, stumpfe_exploring_2012,stumpfe_recent_2014,stumpfe_evolving_2019} can pose challenges in downstream modeling tasks depending on the choice of molecular representation. As the assumption that similar molecules have similar properties breaks down, a dataset will contain both more and sharper activity cliffs, resulting in a ``rougher'' structure-activity landscape. In turn, these rougher QSPR landscapes make generalization harder due to the increasingly complex relationship between molecular structure and properties.  

Dataset roughness is typically assessed qualitatively, although there are metrics that attempt to quantify this, such as the structure-activity relationship index (SARI) \autocite{peltason_sar_2007} and the modelability index (MODI) \autocite{golbraikh_data_2014}. However, these metrics are primarily intended for application to bioactivity datasets (SARI) or to classification datasets (MODI), and extending these metrics to arbitrary regression tasks remains a challenge. To address this, we have recently proposed the \rogiOrig{} (ROGI) \autocite{aldeghi_roughness_2022}, a scalar metric that captures global surface roughness by measuring the loss in the dispersion of molecular properties as a dataset is progressively coarse-grained. Briefly, we are given an input representation for each molecule $x \in \mathbb{R}^d$ and a distance metric $d : \mathbb{R}^d \times \mathbb{R}^d \rightarrow \mathbb{R}$, then (1) the dataset is clustered using complete linkage clustering at a given distance threshold $t$, (2) the dataset is coarse-grained by replacing the property label $y_i$ of each point with the mean of its respective cluster $\bar{y}_j$, (3) the standard deviation of the coarse-grained dataset $\sigma_t$ is calculated,  (4) steps (1)--(3) are repeated for $t \in [0, \ldots, \max d_x]$, (5) the area under the curve of $2\,(\sigma_0 - \sigma_t)$ vs. $t$ is measured to yield the ROGI. Datasets with larger ROGI values result in larger cross-validated model errors, consistent with intuition. Across a variety of datasets from GuacaMol\autocite{brown_guacamol_2019}, TDC\autocite{huang_artificial_2022}, and ChEMBL\autocite{gaulton_chembl_2012} and machine learning (ML) model architectures, the ROGI correlates strongly with cross-validated model root-mean-square error (RMSE) and generally outperforms alternative metrics \autocite{aldeghi_roughness_2022}. 

Given these strong correlations, we sought to broadly examine recent claims about the superiority of molecular representations learned by ``foundation models'' for chemistry\autocite{ahmad_chemberta-2_2022,mendez-lucio_mole_2022,fabian_molecular_2020,rong_self-supervised_2020,wang_smiles-bert_2019,ross_large-scale_2022} through the lens of QSPR surface roughness. Foundation models are a class of ML models that are trained on large, unlabeled datasets via self-supervised learning (sometimes supervised learning) and are in principle capable of adapting rapidly to downstream tasks with very few labeled data points \autocite{bommasani_opportunities_2022}. Pretrained foundation models are now standard practice in several domains, such as natural language processing\autocite{devlin_bert_2019,raffel_exploring_2020,brown_language_2020}, computer vision\autocite{yuan_florence_2021,radford_learning_2021}, and protein modeling\autocite{brandes_proteinbert_2022,lin_evolutionary-scale_2023}. Given the abundance of unlabeled chemical data and the limited amount of data encountered in many property prediction tasks, foundation models may benefit chemistry by learning meaningful molecular representations suitable for property prediction tasks in the low data regime.

Despite this interest, empirical evaluation of proposed chemical foundation models has shown mixed results. Recent work from \tcite{deng_taking_2022} assessed the performance both SMILES- and graph-based pretrained chemical models (PCMs), MolBERT\autocite{fabian_molecular_2020} and GROVER\autocite{rong_self-supervised_2020}, respectively, on a variety of benchmark tasks from MoleculeNet\autocite{wu_moleculenet_2018} and opioid bioactivity datasets from ChEMBL\autocite{gaulton_chembl_2012}. For each task, they compared the performance of these proposed chemical foundation models to a random forest model trained on radius 2, 2048-bit Morgan fingerprints. The authors found that this baseline was competitive for many benchmark tasks and even superior in several of the opioid tasks. This finding is consistent with results reported in the PCM literature where learned representations offer inconsistent improvement over baseline approaches.

In this work, we complement this analysis by characterizing the roughness of the QSPR surfaces generated by PCMs on both toy and experimental modeling tasks. To do so, we reformulate ROGI as \newrogi{} to enable cross-representation comparison. While the original ROGI correlates strongly with cross-validated RMSE across datasets when holding the representation constant, it does not necessarily provide a meaningful basis for comparison among representations due to the relationship between distances and the dimensionality of a given representation. We show that for a variety of PCMs (VAE \autocite{gomez-bombarelli_automatic_2018}, GIN\autocite{xu_how_2019,hu_strategies_2020}, ChemBERTa\autocite{ahmad_chemberta-2_2022}, and ChemGPT\autocite{frey_neural_2022}) and a variety of molecular tasks, learned molecular representations do not provide smoother structure-property relationships than simple descriptor and fingerprint representations. The failure of PCMs to learn a continuous embedding of molecular structures that smoothly correlates with various properties of interest both (a) explains their poor empirical performance in property prediction tasks without fine-tuning and (b) motivates the use of \newrogi{} to evaluate smoothness when new pretraining strategies are proposed.

\section*{Results and discussion}

\subsection*{Reformulation of the ROGI as \newrogi{} enables cross-dimensional comparisons}

\begin{figure}[h]
    \centering
    \includegraphics[width=0.9\textwidth]{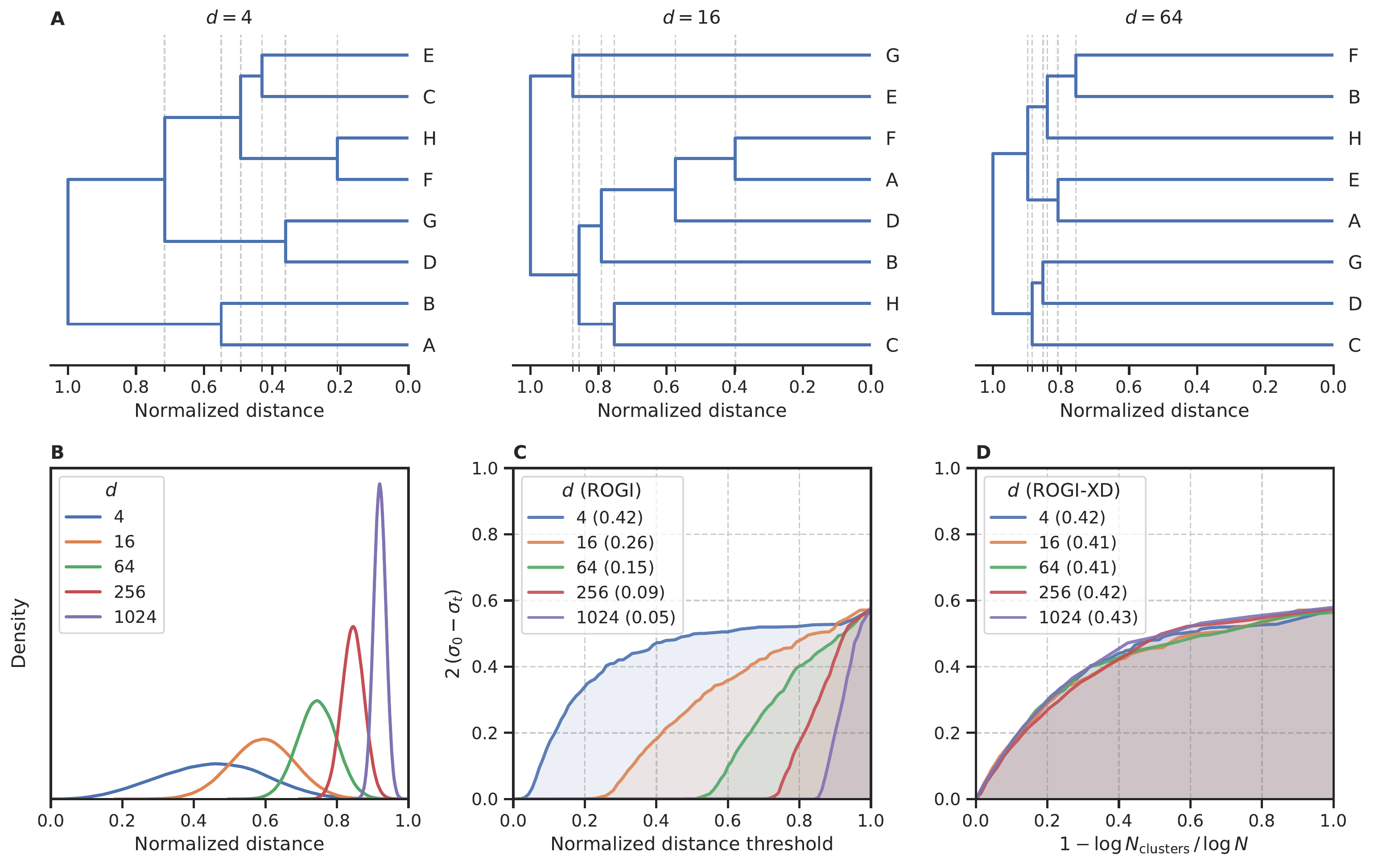}
    \caption{
        Reformulation of ROGI as \newrogi{} enables cross-representation and cross-dimension comparisons.
        \textbf{(A)} The dendrograms produced by complete linkage clustering of eight points sampled uniformly from the domain $\range{0}{1}^d$. As the size of the domain increases, the clustering steps become more compressed and happen closer to 1 (normalized distance). The minor ticks and vertical gridlines in each subplot correspond to a step in the dendrogram.
        \textbf{(B)} The distribution of normalized distance for \num{1000} points drawn from the domain $\range{0}{1}^d$. As the dimensionality increases, the distance distribution sharpens and centers closer to 1.
        \textbf{(C)} Higher dimensional representations produce lower ROGI values.
        \textbf{(D)} Redefining the coarse-graining domain to $1 - \log N_{\mathrm{clusters}} / \log N$ results in similar ROGI values regardless of representation size.}
    \label{fig:dendrogram+toy}
\end{figure}

In its original formulation, ROGI values are not comparable across representations of different dimensionality. Distances between randomly-sampled points generally increase with representation size, so even when normalizing distances to the same range (e.g., $[0, 1]$), higher-dimensional representations do not coarse-grain until larger values of normalized distanced threshold (\autoref{fig:dendrogram+toy}A). Ultimately, this will result in artificially low ROGI values for QSPR datasets with high-dimensional representations. To illustrate this, consider $N$ points sampled uniformly from the unit hypercube of dimension $d$ with random property labels $y \sim \mathcal{U}(0, 1)$. As $|d|$ increases, the normalized distance distribution of these points will become more tightly peaked and centered closer to 1 (\autoref{fig:dendrogram+toy}B), which results in the delayed coarse-graining phenomenon mentioned earlier. This delayed coarse-graining causes the curve of loss of dispersion $2\,(\sigma_0 - \sigma_t)$ vs. normalized distance threshold $t$ to be depressed at lower values of $t$, producing lower ROGI values for higher dimensional representations (\autoref{fig:dendrogram+toy}C). It could be argued that a higher-dimensional representation may result in a ``smoother'' representation due to the larger distances between points, but for large differences in $|d|$, the ROGI essentially becomes a proxy for the inverse of representation size rather than differences in the underlying SPR surface. The datasets sampled from the unit hypercube abstractly represent the ``same'' dataset in hyperspace as $N \rightarrow \infty$, so they should possess roughly equal roughness values when controlling for $|d|$.

To minimize the impact of dimensionality on the ROGI, we change its integration variable to capture the degree of coarse-graining independent of representation size.
Procedurally, ``coarse-graining'' entails taking a step up in the dendrogram produced during the clustering routine. Whereas originally we scan along the distance required to take such a step, we now opt to use $1 - \log N_{\mathrm{clusters}} / \log N$, where $N_{\mathrm{clusters}}$ is the number of clusters at the given step in the dendrogram and $N$ is the dataset size. This new formulation, which we refer to as \newrogi{}, produces similar values for each toy dataset regardless of its dimensionality (\autoref{fig:dendrogram+toy}D). We note that while there are other formulations that reflect a similar concept, they must possess a constant integration domain. For example, using $1 - N_{\mathrm{clusters}} / N$ as the $x$-axis produces a similar trend as above (\autoref{fig:reformulation-ex}), but it is defined on the domain $\range{0}{1 - 1/N}$, thus making the score dependent on $N$ and confounding comparisons across datasets with large differences in size.

\subsection*{The \newrogi{} correlates strongly across representations}
We next sought to evaluate how well the \newrogi{} correlates with model error across chemical representations. First, we measured \newrogi{} and cross-validated RMSE for all combinations of task, model, and representation then calculated the Pearson correlation coefficient $r$ between \newrogi{} and cross-validated RMSE across representations for the \emph{same} task and model. We analyze a variety of regression tasks using experimental ADMET datasets from the TDC \autocite{huang_artificial_2022} and datasets generated using GuacaMol \autocite{brown_guacamol_2019} oracle functions, and we use the same ML models as in our previous study.\autocite{aldeghi_roughness_2022} We look at two fixed representations: molecular descriptors and Morgan fingerprints; four pretrained representations: SMILES \vae{} (VAE) \autocite{gomez-bombarelli_automatic_2018}, \gin{} (GIN) \autocite{xu_how_2019} pretrained with node attribute masking \autocite{hu_strategies_2020}, ChemBERTa \autocite{ahmad_chemberta-2_2022}, and ChemGPT \autocite{frey_neural_2022}; and 128-dimensional random embeddings. For more details, see the \nameref{sec:methods} section below.

\begin{figure}[t!]
    \centering
    \includegraphics[width=0.9\textwidth]{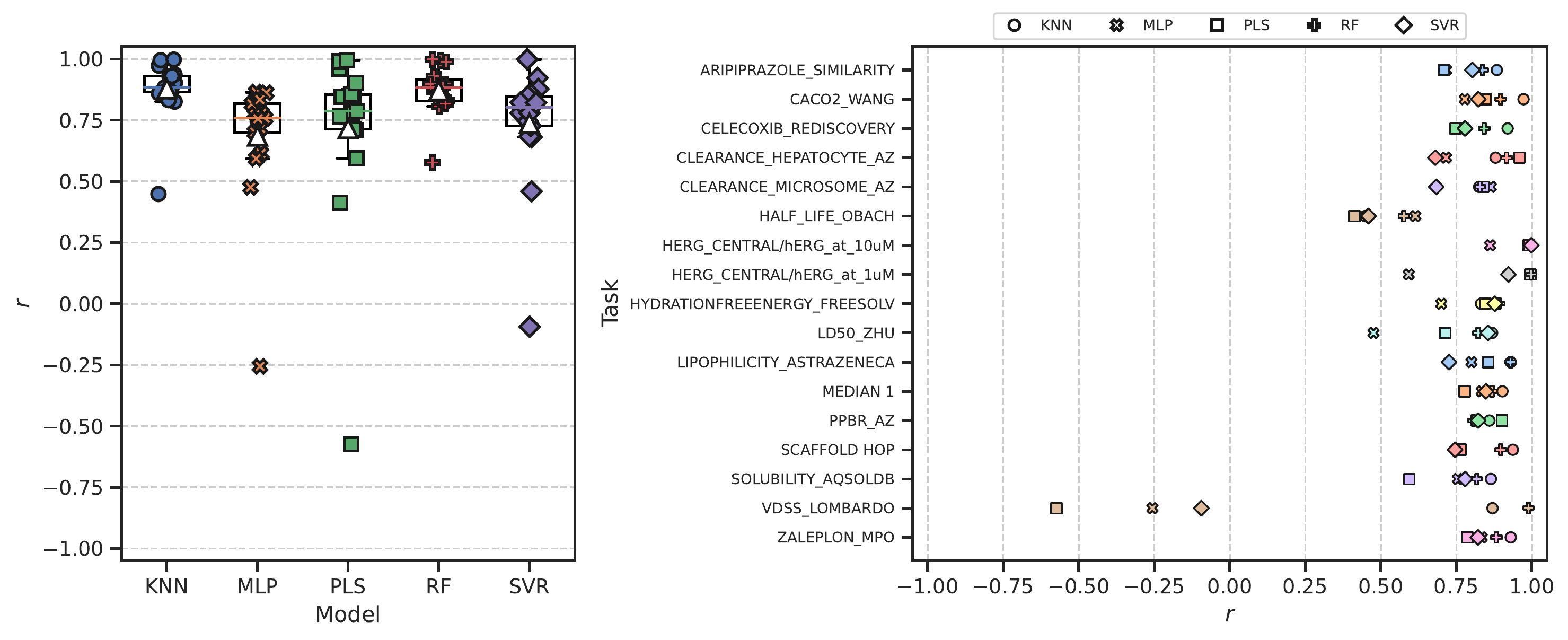}
    \caption{Distribution of Pearson correlation coefficients $r$ between \newrogi{} and cross-validated RMSE across all representations evaluated for a given ML model and task. \textit{Left}: Box plot of correlations grouped by ML model architecture with individual data points plotted above. The median is depicted via the solid, colored line, and the mean by the white triangle ($\triangle$). \textit{Right}: Correlations grouped by task. \modelCaption.}
    \label{fig:corr-model-task}
\end{figure}

The \newrogi{} produces strong correlations with model error across molecular representation for the  majority of tasks and ML models tested (\autoref{fig:corr-model-task}). The median correlation across all combinations of model and task ranges between 0.72 and 0.88, with the best correlations observed for both the \rf{} (RF) and \knn{} (KNN) models. This is in contrast to the original ROGI, which generally produces weak correlations (median $r \in \range{-0.32}{0.28}$) when subjected to the same analysis (\autoref{fig:model-task-v1}). As shown in the toy example above, the original ROGI is affected by representation size, so the range of dimensionalities in the representations tested (14 to 2048, \autoref{tbl:repr-sizes}) negatively impacts correlation strength.

We note that when we measure the correlation between \newrogi{} and RMSE across tasks for a given model using molecular descriptors (as in our original study), we see similarly strong correlations (\autoref{fig:tasks-both}). These correlations remain strong when we measure correlation over \emph{both} representations and tasks, whereas they decrease significantly with the original ROGI (\autoref{tbl:corr-task}). In turn, this allows for the direct comparison of ROGI values measured for two datasets with differing representations.

\begin{table}[h]
\centering
\begin{threeparttable}
\caption{Pearson correlation coefficient $r$ between roughness metric and cross-validated RMSE across all tasks and representations for a given model.}
\label{tbl:corr-task}
    \begin{tabular}{@{}crrrrr@{}}
    \toprule
     & \multicolumn{5}{c}{model} \\ \cmidrule(l){2-6} 
    metric & \multicolumn{1}{l}{KNN} & \multicolumn{1}{l}{MLP} & \multicolumn{1}{l}{PLS} & \multicolumn{1}{l}{RF} & \multicolumn{1}{l}{SVR} \\ \midrule
    ROGI & 0.800 & 0.675 & 0.835 & 0.809 & 0.771 \\
    \newrogi{} & \textbf{0.990} & \textbf{0.913} & \textbf{0.983} & \textbf{0.985} & \textbf{0.958} \\ \bottomrule
    \end{tabular}
\end{threeparttable}
\end{table}

\begin{figure}[t!]
    \centering
    \includegraphics[width=0.25\textwidth]{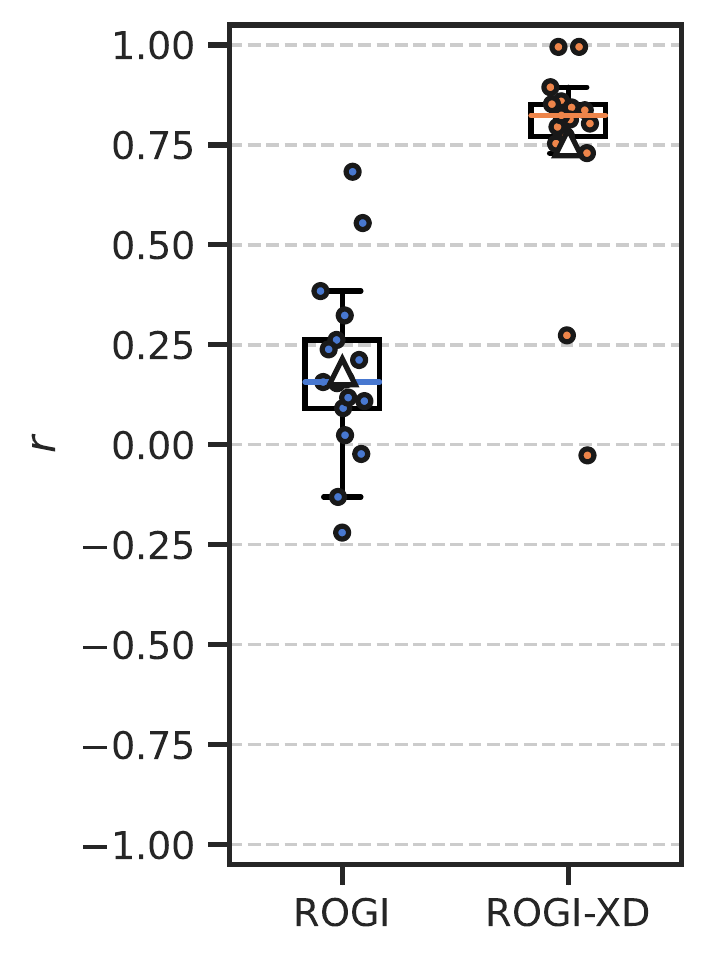}
    \caption{Distribution of Pearson correlation coefficients $r$ between roughness metric and minimum model error for a given task across all representations. Each point corresponds to an individual task. The median is depicted via the solid, colored line, and the mean by the white triangle ($\triangle$).}
    \label{fig:corr-model-best}
\end{figure}

It is also possible to measure the correlation between ROGI or \newrogi{} and the \emph{minimum} model error for a given task and representation. In other words, rather than treating each ML model separately as above, we now (1) measure the model error and roughness metric for all combinations of task, representation, and ML model; (2) take the minimum model error for each combination of task and representation; and (3) measure the correlation between the roughness metric and this ``best-case'' model error for each task. The \newrogi{} again produces strong correlations across all datasets (median $r=0.82$) compared to the original ROGI (median $r=0.16$) (\autoref{fig:corr-model-best}). This discrepancy in correlation strength is expected because  this analysis still relies on comparisons across representations. The \newrogi{}'s strong correlation with best model error across representations thus allows a user to quickly get an idea of best-case model performance for a variety of representations without resorting to empirical testing. This can further be extended to comparing best-case modelability among datasets given a set of possible representations by calculating the \newrogi{} for each representation and then selecting the lowest one for the task.

\subsection*{Pretrained molecular representations do not provide smoother structure-activity landscapes than fingerprints and descriptors}

\begin{figure}[b!]
    \centering
    \includegraphics[width=0.9\textwidth]{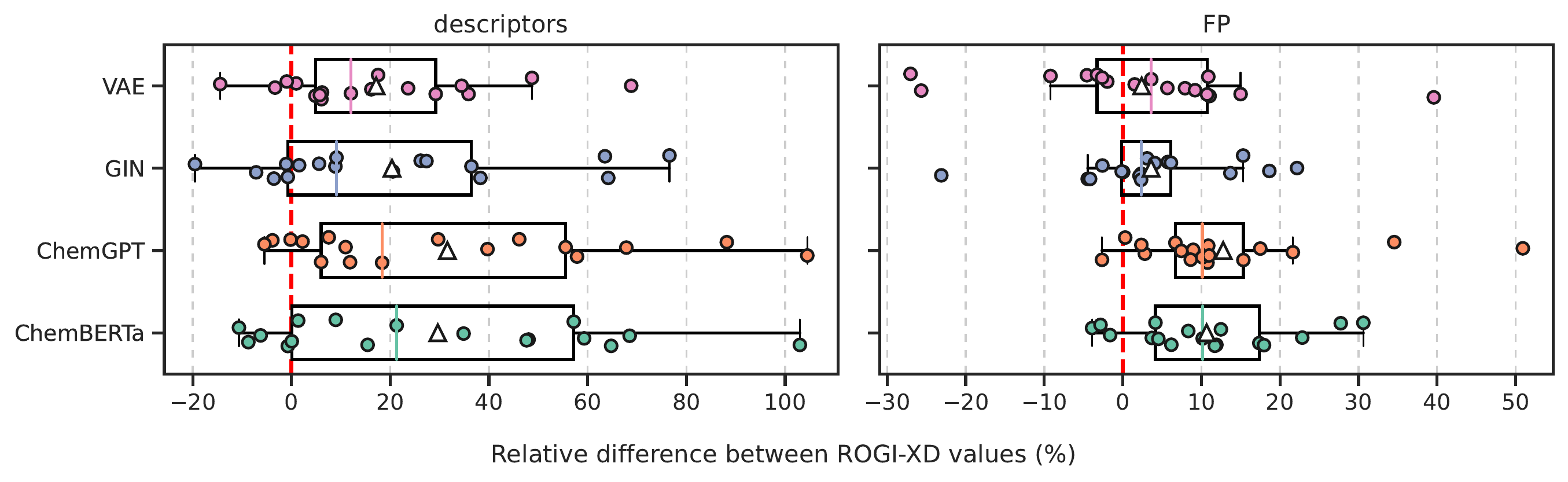}
    \caption{Distribution of relative difference between \newrogi{} values of the given pretrained representation ($y$-axis) and fixed representation (top) for each task. Positive values indicate that the pretrained representation produces a rougher QSPR surface, and negative values indicate the opposite. Individual data are plotted above and box plot is plotted above. The median is depicted via the solid, colored line, and the mean by the white triangle ($\triangle$). \reprCaption.}
    \label{fig:rogi-boxplot}
\end{figure}

The \newrogi{} formulation's strong correlation with model error across representations thus allows us to compare the smoothness of learned representations to that of fixed representations. We use the same pretrained representations as before to broadly survey different strategies of learning molecular representations: a recurrent neural network-based encoder-decoder framework (VAE), graph-based pretraining (GIN), encoder-only large language model (LLM; ChemBERTa), and decoder-only LLM (ChemGPT). For each task, we calculate the relative difference between \newrogi{} values for each pair of pretrained and fixed representations (i.e., $\text{\newrogi{}}_p \,/\, \text{\newrogi{}}_f - 1$, where $\text{\newrogi{}}_p$ and $\text{\newrogi{}}_f$ are the \newrogi{} values of a given pretrained and fixed representation, respectively, for the same task). While there are other pretraining techniques and model architectures available, we do not intend for this analysis to be exhaustive. Rather, our goal is to provide a supplementary technique by which to understand trends in model performance.

We find that across all tasks tested above, PCMs do not generate quantitatively smoother QSPR surfaces when compared to those generated via molecular descriptors or fingerprints (\autoref{fig:rogi-boxplot}). In more than 50\% of the tasks evaluated, both descriptors and fingerprints generated smoother QSPR surfaces. The median relative \newrogi{} values for each pretrained representation compared to descriptors and fingerprints range between 9.1--21.3\% and 2.3--10.1\%, respectively. Indeed, these \newrogi{} values are consistent with the cross-validation results of descriptors and fingerprints being generally lower in RMSE than the pretrained representations (\Cref{fig:all-v2-1,fig:all-v2-2}). An extreme case is the \texttt{Scaffold Hop} task, where the GIN, ChemGPT, and ChemBERTa representations produce \newrogi{}s of 0.150, 0.174, and 0.172, respectively, compared to the 0.085 of descriptors. However, we emphasize that PCMs do not generate \emph{bad} representations, but rather that these learned representations are \emph{not smoother} than simple, fixed representations.

Studies that introduce new pretraining techniques or model architectures rarely, if ever, analyze the smoothness of the resulting QSPR surfaces. Rather, they benchmark their method on a variety of property prediction tasks and frequently report mixed results; on some tasks, the new technique outperforms the current state-of-the-art, but on others, it fails to compete with simple baselines. We believe that the lack of smoothness at least partially explains their inability to consistently outperform established molecular representations on supervised learning benchmarks.

\section*{Conclusion}
We have described \newrogi{}, a reformulation of the ROGI that enables comparison of structure-activity roughness values across representations via changing the integration variable. The \newrogi{} correlates strongly with cross-validated model error both across representations for a given task (median $r = 0.72\text{--}0.88$) and over both representations and tasks (median $r = 0.91\text{--}0.99$). We then use \newrogi{} to evaluate several recently-reported pretrained chemical representations from SMILES VAE, GIN, ChemGPT, and ChemBERTa. Across all of the tasks evaluated, the \newrogi{} values of these  representations were no smoother than those of fingerprints and descriptors. These results are consistent with the empirical results of various benchmark studies which show that pretrained models are not universally superior to fingerprints or descriptors. 

Taken together, these observations suggest that more work remains in developing chemical foundation models. Though it is unreasonable to expect that any single pretrained representation will produce a smoother QSPR surface in every task, a reasonable desideratum is that such a representation is of comparable smoothness to simple baseline representations for a majority of useful properties. The \newrogi{} is thematically similar to a contrastive loss, as both will scale proportionally with the frequency and severity of activity cliffs in a given dataset. Imposing stronger assumptions of smoothness with respect to molecular structure during model pretraining by weak supervision on simple, calculable properties could aid in producing smoother QSPR surfaces.

A limitation of our analysis is that we have treated the pretrained representations as static for downstream modeling; an alternative is to fine-tune them by training the model on additional, labeled data, in turn helping to smooth the corresponding QSPR surface. In a sense, the evaluations here have demonstrated the need for fine-tuning in the absence of a universally smooth representation. This introduces many additional design choices, so we leave this evaluation for future work.

\section*{Materials and methods}\label{sec:methods}

All code and data used in this work is available at: \url{\therepo}

\subsection*{Representations}\label{sec:reps}
\begin{sloppypar}
\paragraph{Descriptors} As in our previous study\autocite{aldeghi_roughness_2022}, we calculate the following 14 molecular descriptors for each molecule using RDKit \autocite{noauthor_rdkit_nodate} and concatenate them to form a vector: \texttt{MolWt}, \texttt{FractionCSP3}, \texttt{NumHAcceptors}, \texttt{NumHDonors}, \texttt{NOCount}, \texttt{NHOHCount}, \texttt{NumAliphaticRings},
\texttt{NumAliphaticHeterocycles}, \texttt{NumAromaticHeterocycles}, \texttt{NumAromaticRings}, \texttt{NumRotatableBonds}, \texttt{TPSA}, \texttt{qed}, and \texttt{MolLogP}. After the representation was calculated for each molecule in a dataset, each feature axis was scaled to the range $[0, 1]$.
\end{sloppypar}

\paragraph{Fingerprints} Morgan fingerprints of radius 2 and 512 bits were calculated via RDKit.\autocite{noauthor_rdkit_nodate} We note that this differs from our prior study, which used 2048-bit Morgan fingerprints, because most datasets evaluated contained on the order of \num{1000} datapoints. We did not observe a significant difference in performance as a function of fingerprint length.

\paragraph{VAE} We implement a character-based \vae{} of SMILES strings based on the architecture of \tcite{gomez-bombarelli_automatic_2018} using the tokenization scheme of \tcite{schwaller_mapping_2021} The model was trained on the ZINC 250k dataset using an 80/20 training/validation split for 100 epochs and early stopping tracking the validation loss with a patience of five epochs. The 128-dimensional mean latent codes produced by the encoder were used as the molecular representations.

\paragraph{GIN} We pretrain a \gin{} \autocite{xu_how_2019} on molecular graphs via node attribute masking \autocite{hu_strategies_2020} using the TorchDrug \autocite{zhu_torchdrug_2022} implementation. The model is trained on the ZINC 250k dataset using a mask rate of 15\%, batch normalization, 80/20 training/validation split for 100 epochs, and early stopping tracking the validation loss with a patience of five epochs. The 300-dimensional molecule-level representation is generated via the mean of all node-level representations from the final iteration of message-passing.

\paragraph{ChemBERTa} We used the \texttt{ChemBERTa-77M-MLM} model available from the Hugging Face hub \autocite{noauthor_deepchemchemberta-77m-mlm_nodate}. ChemBERTa is a RoBERTa-style model pretrained on SMILES strings using 77M molecules from PubChem using masked-language modeling originally reported in \tcite{ahmad_chemberta-2_2022}. The 384-dimensional embeddings of the \texttt{[CLS]} token in a sequence from the final transformer layer were taken as the molecule-level representation. Batches of sequences were padded right-wise.

\paragraph{ChemGPT} We used the \texttt{ChemGPT-1.2B} model available from the Hugging Face hub \autocite{noauthor_ncfreychemgpt-12b_nodate}. ChemGPT is a GPT-style model pretrained on SELFIES \autocite{krenn_self-referencing_2020} strings from the PubChem 10M dataset \autocite{chithrananda_chemberta_2020} using causal language modeling originally reported in \tcite{frey_neural_2022}. GPT-style models are decoder-only transformers, so molecule-level representations were taken as the 2048-dimensional embeddings of the \textit{right-most} token in a sequence from the final transformer layer. Given this right-wise embedding scheme, batches of sequences were padded \textit{left-wise}.

\paragraph{Random} Embeddings were uniformly sampled from the domain $[0, 1]^{128}$.

\subsection*{Tasks} As in our previous work \autocite{aldeghi_roughness_2022}, we used two groups of tasks, (1) ADME and toxicity datasets from the TDC \autocite{huang_artificial_2022} and (2) toy datasets generated by sampling \num{10000} molecules from the ZINC250k dataset and then calculating the following GuacaMol \autocite{brown_guacamol_2019} oracle function values for these molecules: \texttt{Scaffold Hop}, \texttt{Median 1}, \texttt{Aripiprazole\_Similarity}, \texttt{Zaleplon\_MPO}, \texttt{Celecoxib\_Rediscovery}. We exclude many of the original GuacaMol tasks, as their oracle functions use descriptor values in the scoring function that overlap with our descriptor representation. For the \texttt{hERG\_at\_1uM} and \texttt{hERG\_at\_10uM} tasks from the TDC, datasets were downsampled to \num{10000} molecules; in these instances, reported ROGI values are the mean of five random subsamples.

\subsection*{Cross-validation} As in previous work \autocite{aldeghi_roughness_2022}, we performed 5-fold cross-validation using the following five regression models from Scikit-learn \autocite{pedregosa_scikit-learn_2011}: \knn{} (KNN), and \mlp{} (MLP), \pls{} (PLS), \rf{} (RF), and \svr{} (SVR). All models utilized default settings except for the RF model, which used an \texttt{n\_estimators} value of 50. We scale the property labels to the range $[0, 1]$ before cross-validation and report the mean root-mean-squared error (RMSE) of all five folds.

\section*{Acknowledgements}
This work was funded by the MIT-IBM Watson AI Lab. The authors thank Jenna Fromer and Itai Levin for commenting on the manuscript.

\printbibliography

@book{silipo_qsar_1991,
	address = {Amsterdam; New York},
	title = {{QSAR}: rational approaches to the design of bioactive compounds: Proceedings of the {VIII} {European} {Symposium} on {Quantitative} {Structure}-{Activity} {Relationships}, {Sorrento}, {Italy}, 9-13 {September} 1990},
	isbn = {978-0-444-88839-6},
	shorttitle = {{QSAR}, rational approaches to the design of bioactive compounds},
	language = {eng},
	publisher = {Elsevier Science},
	editor = {Silipo, Carlo and Vittoria, Antonio},
	year = {1991},
}

@article{maggiora_outliers_2006,
	title = {On {Outliers} and {Activity} {CliffsWhy} {QSAR} {Often} {Disappoints}},
	volume = {46},
	issn = {1549-9596},
	url = {https://doi.org/10.1021/ci060117s},
	doi = {10.1021/ci060117s},
	number = {4},
	urldate = {2023-04-25},
	journal = {Journal of Chemical Information and Modeling},
	author = {Maggiora, Gerald M.},
	month = jul,
	year = {2006},
	pages = {1535--1535},
}

@article{stumpfe_exploring_2012,
	title = {Exploring {Activity} {Cliffs} in {Medicinal} {Chemistry}},
	volume = {55},
	issn = {0022-2623},
	url = {https://doi.org/10.1021/jm201706b},
	doi = {10.1021/jm201706b},
	number = {7},
	urldate = {2023-04-25},
	journal = {Journal of Medicinal Chemistry},
	author = {Stumpfe, Dagmar and Bajorath, Jürgen},
	month = apr,
	year = {2012},
	pages = {2932--2942},
}

@article{stumpfe_recent_2014,
	title = {Recent {Progress} in {Understanding} {Activity} {Cliffs} and {Their} {Utility} in {Medicinal} {Chemistry}},
	volume = {57},
	issn = {0022-2623},
	url = {https://doi.org/10.1021/jm401120g},
	doi = {10.1021/jm401120g},
	number = {1},
	urldate = {2023-04-25},
	journal = {Journal of Medicinal Chemistry},
	author = {Stumpfe, Dagmar and Hu, Ye and Dimova, Dilyana and Bajorath, Jürgen},
	month = jan,
	year = {2014},
	pages = {18--28},
}

@article{stumpfe_evolving_2019,
	title = {Evolving {Concept} of {Activity} {Cliffs}},
	volume = {4},
	url = {https://doi.org/10.1021/acsomega.9b02221},
	doi = {10.1021/acsomega.9b02221},
	number = {11},
	urldate = {2023-04-25},
	journal = {ACS Omega},
	author = {Stumpfe, Dagmar and Hu, Huabin and Bajorath, Jürgen},
	month = sep,
	year = {2019},
	pages = {14360--14368},
}

@article{ross_large-scale_2022,
	title = {Large-scale chemical language representations capture molecular structure and properties},
	volume = {4},
	issn = {2522-5839},
	url = {https://www.nature.com/articles/s42256-022-00580-7},
	doi = {10.1038/s42256-022-00580-7},
	language = {en},
	number = {12},
	urldate = {2023-04-25},
	journal = {Nature Machine Intelligence},
	author = {Ross, Jerret and Belgodere, Brian and Chenthamarakshan, Vijil and Padhi, Inkit and Mroueh, Youssef and Das, Payel},
	month = dec,
	year = {2022},
	pages = {1256--1264},
}

@article{gaulton_chembl_2012,
	title = {{ChEMBL}: a large-scale bioactivity database for drug discovery},
	volume = {40},
	issn = {0305-1048},
	shorttitle = {{ChEMBL}},
	url = {https://www.ncbi.nlm.nih.gov/pmc/articles/PMC3245175/},
	doi = {10.1093/nar/gkr777},
	number = {Database issue},
	urldate = {2023-04-24},
	journal = {Nucleic Acids Research},
	author = {Gaulton, Anna and Bellis, Louisa J. and Bento, A. Patricia and Chambers, Jon and Davies, Mark and Hersey, Anne and Light, Yvonne and McGlinchey, Shaun and Michalovich, David and Al-Lazikani, Bissan and Overington, John P.},
	month = jan,
	year = {2012},
	pmid = {21948594},
	pmcid = {PMC3245175},
	pages = {D1100--D1107},
}

@article{wu_moleculenet_2018,
	title = {{MoleculeNet}: a benchmark for molecular machine learning},
	volume = {9},
	issn = {2041-6539},
	shorttitle = {{MoleculeNet}},
	url = {https://pubs.rsc.org/en/content/articlelanding/2018/sc/c7sc02664a},
	doi = {10.1039/C7SC02664A},
	language = {en},
	number = {2},
	urldate = {2023-04-24},
	journal = {Chemical Science},
	author = {Wu, Zhenqin and Ramsundar, Bharath and Feinberg, Evan N. and Gomes, Joseph and Geniesse, Caleb and Pappu, Aneesh S. and Leswing, Karl and Pande, Vijay},
	month = jan,
	year = {2018},
	pages = {513--530},
}

@article{deng_taking_2022,
	title = {Taking a {Respite} from {Representation} {Learning} for {Molecular} {Property} {Prediction}},
	url = {http://arxiv.org/abs/2209.13492},
	urldate = {2023-03-31},
	author = {Deng, Jianyuan and Yang, Zhibo and Wang, Hehe and Ojima, Iwao and Samaras, Dimitris and Wang, Fusheng},
	month = oct,
	year = {2022},
	journal = {arXiv:2209.13492 [cs, q-bio]},
}

@article{mendez-lucio_mole_2022,
	title = {{MolE}: a molecular foundation model for drug discovery},
	shorttitle = {{MolE}},
	url = {http://arxiv.org/abs/2211.02657},
	urldate = {2023-03-30},
	author = {Méndez-Lucio, Oscar and Nicolaou, Christos and Earnshaw, Berton},
	month = nov,
	year = {2022},
	journal = {arXiv:2211.02657 [cs, q-bio]},
}

@inproceedings{wang_smiles-bert_2019,
	address = {Niagara Falls, NY, USA},
	title = {{SMILES}-{BERT}: {Large} {Scale} {Unsupervised} {Pre}-{Training} for {Molecular} {Property} {Prediction}},
	isbn = {978-1-4503-6666-3},
	shorttitle = {{SMILES}-{BERT}},
	url = {https://dl.acm.org/doi/10.1145/3307339.3342186},
	doi = {10.1145/3307339.3342186},
	language = {en},
	urldate = {2023-03-30},
	booktitle = {Proceedings of the 10th {ACM} {International} {Conference} on {Bioinformatics}, {Computational} {Biology} and {Health} {Informatics}},
	publisher = {ACM},
	author = {Wang, Sheng and Guo, Yuzhi and Wang, Yuhong and Sun, Hongmao and Huang, Junzhou},
	month = sep,
	year = {2019},
	pages = {429--436},
}

@article{fabian_molecular_2020,
	title = {Molecular representation learning with language models and domain-relevant auxiliary tasks},
	url = {http://arxiv.org/abs/2011.13230},
	urldate = {2023-04-10},
	author = {Fabian, Benedek and Edlich, Thomas and Gaspar, Héléna and Segler, Marwin and Meyers, Joshua and Fiscato, Marco and Ahmed, Mohamed},
	month = nov,
	year = {2020},
	journal = {arXiv:2011.13230 [cs]},
}

@article{rong_self-supervised_2020,
	title = {Self-{Supervised} {Graph} {Transformer} on {Large}-{Scale} {Molecular} {Data}},
	url = {http://arxiv.org/abs/2007.02835},
	urldate = {2023-04-24},
	author = {Rong, Yu and Bian, Yatao and Xu, Tingyang and Xie, Weiyang and Wei, Ying and Huang, Wenbing and Huang, Junzhou},
	month = oct,
	year = {2020},
	journal = {arXiv:2007.02835 [cs, q-bio]},
}

@article{bommasani_opportunities_2022,
	title = {On the {Opportunities} and {Risks} of {Foundation} {Models}},
	url = {http://arxiv.org/abs/2108.07258},
	doi = {10.48550/arXiv.2108.07258},
	urldate = {2023-04-24},
	author = {Bommasani, Rishi and Hudson, Drew A. and Adeli, Ehsan and Altman, Russ and Arora, Simran and von Arx, Sydney and Bernstein, Michael S. and Bohg, Jeannette and Bosselut, Antoine and Brunskill, Emma and Brynjolfsson, Erik and Buch, Shyamal and Card, Dallas and Castellon, Rodrigo and Chatterji, Niladri and Chen, Annie and Creel, Kathleen and Davis, Jared Quincy and Demszky, Dora and Donahue, Chris and Doumbouya, Moussa and Durmus, Esin and Ermon, Stefano and Etchemendy, John and Ethayarajh, Kawin and Fei-Fei, Li and Finn, Chelsea and Gale, Trevor and Gillespie, Lauren and Goel, Karan and Goodman, Noah and Grossman, Shelby and Guha, Neel and Hashimoto, Tatsunori and Henderson, Peter and Hewitt, John and Ho, Daniel E. and Hong, Jenny and Hsu, Kyle and Huang, Jing and Icard, Thomas and Jain, Saahil and Jurafsky, Dan and Kalluri, Pratyusha and Karamcheti, Siddharth and Keeling, Geoff and Khani, Fereshte and Khattab, Omar and Koh, Pang Wei and Krass, Mark and Krishna, Ranjay and Kuditipudi, Rohith and Kumar, Ananya and Ladhak, Faisal and Lee, Mina and Lee, Tony and Leskovec, Jure and Levent, Isabelle and Li, Xiang Lisa and Li, Xuechen and Ma, Tengyu and Malik, Ali and Manning, Christopher D. and Mirchandani, Suvir and Mitchell, Eric and Munyikwa, Zanele and Nair, Suraj and Narayan, Avanika and Narayanan, Deepak and Newman, Ben and Nie, Allen and Niebles, Juan Carlos and Nilforoshan, Hamed and Nyarko, Julian and Ogut, Giray and Orr, Laurel and Papadimitriou, Isabel and Park, Joon Sung and Piech, Chris and Portelance, Eva and Potts, Christopher and Raghunathan, Aditi and Reich, Rob and Ren, Hongyu and Rong, Frieda and Roohani, Yusuf and Ruiz, Camilo and Ryan, Jack and Ré, Christopher and Sadigh, Dorsa and Sagawa, Shiori and Santhanam, Keshav and Shih, Andy and Srinivasan, Krishnan and Tamkin, Alex and Taori, Rohan and Thomas, Armin W. and Tramèr, Florian and Wang, Rose E. and Wang, William and Wu, Bohan and Wu, Jiajun and Wu, Yuhuai and Xie, Sang Michael and Yasunaga, Michihiro and You, Jiaxuan and Zaharia, Matei and Zhang, Michael and Zhang, Tianyi and Zhang, Xikun and Zhang, Yuhui and Zheng, Lucia and Zhou, Kaitlyn and Liang, Percy},
	month = jul,
	year = {2022},
	journal = {arXiv:2108.07258 [cs]},
}

@article{brandes_proteinbert_2022,
	title = {{ProteinBERT}: a universal deep-learning model of protein sequence and function},
	volume = {38},
	issn = {1367-4803},
	shorttitle = {{ProteinBERT}},
	url = {https://doi.org/10.1093/bioinformatics/btac020},
	doi = {10.1093/bioinformatics/btac020},
	number = {8},
	urldate = {2023-04-10},
	journal = {Bioinformatics},
	author = {Brandes, Nadav and Ofer, Dan and Peleg, Yam and Rappoport, Nadav and Linial, Michal},
	month = apr,
	year = {2022},
	pages = {2102--2110},
}

@article{lin_evolutionary-scale_2023,
	title = {Evolutionary-scale prediction of atomic-level protein structure with a language model},
	volume = {379},
	url = {https://www.science.org/doi/10.1126/science.ade2574},
	doi = {10.1126/science.ade2574},
	number = {6637},
	urldate = {2023-04-18},
	journal = {Science},
	author = {Lin, Zeming and Akin, Halil and Rao, Roshan and Hie, Brian and Zhu, Zhongkai and Lu, Wenting and Smetanin, Nikita and Verkuil, Robert and Kabeli, Ori and Shmueli, Yaniv and dos Santos Costa, Allan and Fazel-Zarandi, Maryam and Sercu, Tom and Candido, Salvatore and Rives, Alexander},
	month = mar,
	year = {2023},
	pages = {1123--1130},
}

@article{radford_learning_2021,
	title = {Learning {Transferable} {Visual} {Models} {From} {Natural} {Language} {Supervision}},
	url = {http://arxiv.org/abs/2103.00020},
	doi = {10.48550/arXiv.2103.00020},
	urldate = {2023-04-10},
	author = {Radford, Alec and Kim, Jong Wook and Hallacy, Chris and Ramesh, Aditya and Goh, Gabriel and Agarwal, Sandhini and Sastry, Girish and Askell, Amanda and Mishkin, Pamela and Clark, Jack and Krueger, Gretchen and Sutskever, Ilya},
	month = feb,
	year = {2021},
	journal = {arXiv:2103.00020 [cs]},
}

@article{yuan_florence_2021,
	title = {Florence: {A} {New} {Foundation} {Model} for {Computer} {Vision}},
	shorttitle = {Florence},
	url = {http://arxiv.org/abs/2111.11432},
	doi = {10.48550/arXiv.2111.11432},
	urldate = {2023-04-10},
	author = {Yuan, Lu and Chen, Dongdong and Chen, Yi-Ling and Codella, Noel and Dai, Xiyang and Gao, Jianfeng and Hu, Houdong and Huang, Xuedong and Li, Boxin and Li, Chunyuan and Liu, Ce and Liu, Mengchen and Liu, Zicheng and Lu, Yumao and Shi, Yu and Wang, Lijuan and Wang, Jianfeng and Xiao, Bin and Xiao, Zhen and Yang, Jianwei and Zeng, Michael and Zhou, Luowei and Zhang, Pengchuan},
	month = nov,
	year = {2021},
	journal = {arXiv:2111.11432 [cs]},
}

@article{raffel_exploring_2020,
	title = {Exploring the {Limits} of {Transfer} {Learning} with a {Unified} {Text}-to-{Text} {Transformer}},
	url = {http://arxiv.org/abs/1910.10683},
	doi = {10.48550/arXiv.1910.10683},
	urldate = {2023-04-10},
	author = {Raffel, Colin and Shazeer, Noam and Roberts, Adam and Lee, Katherine and Narang, Sharan and Matena, Michael and Zhou, Yanqi and Li, Wei and Liu, Peter J.},
	month = jul,
	year = {2020},
	journal = {arXiv:1910.10683 [cs, stat]},
}

@article{devlin_bert_2019,
	title = {{BERT}: {Pre}-training of {Deep} {Bidirectional} {Transformers} for {Language} {Understanding}},
	shorttitle = {{BERT}},
	url = {http://arxiv.org/abs/1810.04805},
	doi = {10.48550/arXiv.1810.04805},
	urldate = {2023-04-10},
	author = {Devlin, Jacob and Chang, Ming-Wei and Lee, Kenton and Toutanova, Kristina},
	month = may,
	year = {2019},
	journal = {arXiv:1810.04805 [cs]},
}

@article{brown_language_2020,
	title = {Language {Models} are {Few}-{Shot} {Learners}},
	url = {http://arxiv.org/abs/2005.14165},
	doi = {10.48550/arXiv.2005.14165},
	urldate = {2023-04-24},
	author = {Brown, Tom B. and Mann, Benjamin and Ryder, Nick and Subbiah, Melanie and Kaplan, Jared and Dhariwal, Prafulla and Neelakantan, Arvind and Shyam, Pranav and Sastry, Girish and Askell, Amanda and Agarwal, Sandhini and Herbert-Voss, Ariel and Krueger, Gretchen and Henighan, Tom and Child, Rewon and Ramesh, Aditya and Ziegler, Daniel M. and Wu, Jeffrey and Winter, Clemens and Hesse, Christopher and Chen, Mark and Sigler, Eric and Litwin, Mateusz and Gray, Scott and Chess, Benjamin and Clark, Jack and Berner, Christopher and McCandlish, Sam and Radford, Alec and Sutskever, Ilya and Amodei, Dario},
	month = jul,
	year = {2020},
	journal = {arXiv:2005.14165 [cs]},
}

@article{peltason_sar_2007,
    title = {{SAR} {Index}:  {Quantifying} the {Nature} of {Structure}−{Activity} {Relationships}},
    volume = {50},
    issn = {0022-2623},
    shorttitle = {{SAR} {Index}},
    url = {https://doi.org/10.1021/jm0705713},
    doi = {10.1021/jm0705713},
    number = {23},
    urldate = {2021-05-18},
    journal = {Journal of Medicinal Chemistry},
    author = {Peltason, Lisa and Bajorath, Jürgen},
    month = nov,
    year = {2007},
    pages = {5571--5578},
}

@article{golbraikh_data_2014,
    title = {Data {Set} {Modelability} by {QSAR}},
    volume = {54},
    issn = {1549-9596},
    url = {https://doi.org/10.1021/ci400572x},
    doi = {10.1021/ci400572x},
    number = {1},
    urldate = {2023-03-28},
    journal = {Journal of Chemical Information and Modeling},
    author = {Golbraikh, Alexander and Muratov, Eugene and Fourches, Denis and Tropsha, Alexander},
    month = jan,
    year = {2014},
    pages = {1--4},
}

@online{noauthor_ncfreychemgpt-12b_nodate,
    title = {ncfrey/{ChemGPT}-1.{2B} · {Hugging} {Face}},
    url = {https://huggingface.co/ncfrey/ChemGPT-1.2B},
    urldate = {2023-03-27},
}

@online{noauthor_deepchemchemberta-77m-mlm_nodate,
    title = {{DeepChem}/{ChemBERTa}-{77M}-{MLM} · {Hugging} {Face}},
    url = {https://huggingface.co/DeepChem/ChemBERTa-77M-MLM},
    urldate = {2023-03-27},
}

@article{krenn_self-referencing_2020,
    title = {Self-referencing embedded strings ({SELFIES}): {A} 100\% robust molecular string representation},
    volume = {1},
    issn = {2632-2153},
    shorttitle = {Self-referencing embedded strings ({SELFIES})},
    url = {https://dx.doi.org/10.1088/2632-2153/aba947},
    doi = {10.1088/2632-2153/aba947},
    language = {en},
    number = {4},
    urldate = {2023-03-27},
    journal = {Machine Learning: Science and Technology},
    author = {Krenn, Mario and Häse, Florian and Nigam, AkshatKumar and Friederich, Pascal and Aspuru-Guzik, Alan},
    month = oct,
    year = {2020},
    pages = {045024},
}

@online{noauthor_rdkit_nodate,
    title = {{RDKit}: {Open}-source cheminformatics},
    url = {http://rdkit.org/},
    urldate = {2023-01-05}    
}

@article{gomez-bombarelli_automatic_2018,
    title = {Automatic {Chemical} {Design} {Using} a {Data}-{Driven} {Continuous} {Representation} of {Molecules}},
    volume = {4},
    issn = {2374-7943},
    url = {https://doi.org/10.1021/acscentsci.7b00572},
    doi = {10.1021/acscentsci.7b00572},
    number = {2},
    urldate = {2023-03-20},
    journal = {ACS Cent. Sci.},
    author = {Gómez-Bombarelli, Rafael and Wei, Jennifer N. and Duvenaud, David and Hernández-Lobato, José Miguel and Sánchez-Lengeling, Benjamín and Sheberla, Dennis and Aguilera-Iparraguirre, Jorge and Hirzel, Timothy D. and Adams, Ryan P. and Aspuru-Guzik, Alán},
    month = feb,
    year = {2018},
    pages = {268--276},
    }

@misc{xu_how_2019,
    title = {How {Powerful} are {Graph} {Neural} {Networks}?},
    url = {http://arxiv.org/abs/1810.00826},
    doi = {10.48550/arXiv.1810.00826},
    urldate = {2023-03-20},
    publisher = {arXiv},
    author = {Xu, Keyulu and Hu, Weihua and Leskovec, Jure and Jegelka, Stefanie},
    month = feb,
    year = {2019}  
}

@misc{ahmad_chemberta-2_2022,
    title = {{ChemBERTa}-2: {Towards} {Chemical} {Foundation} {Models}},
    shorttitle = {{ChemBERTa}-2},
    url = {http://arxiv.org/abs/2209.01712},
    doi = {10.48550/arXiv.2209.01712},
    urldate = {2023-03-20},
    publisher = {arXiv},
    author = {Ahmad, Walid and Simon, Elana and Chithrananda, Seyone and Grand, Gabriel and Ramsundar, Bharath},
    month = sep,
    year = {2022}    
}

@article{aldeghi_roughness_2022,
	title = {Roughness of {Molecular} {Property} {Landscapes} and {Its} {Impact} on {Modellability}},
	volume = {62},
	issn = {1549-9596},
	url = {https://doi.org/10.1021/acs.jcim.2c00903},
	doi = {10.1021/acs.jcim.2c00903},
	number = {19},
	urldate = {2023-03-20},
	journal = {Journal of Chemical Information and Modeling},
	author = {Aldeghi, Matteo and Graff, David E. and Frey, Nathan and Morrone, Joseph A. and Pyzer-Knapp, Edward O. and Jordan, Kirk E. and Coley, Connor W.},
	month = oct,
	year = {2022},
	note = {Publisher: American Chemical Society},
	pages = {4660--4671},
}

@article{frey_neural_2022,
    title = {Neural {Scaling} of {Deep} {Chemical} {Models}},
    url = {https://chemrxiv.org/engage/chemrxiv/article-details/627bddd544bdd532395fb4b5},
    doi = {10.26434/chemrxiv-2022-3s512},
    language = {en},
    urldate = {2023-03-20},
    journal = {ChemRxiv},
    author = {Frey, Nathan and Soklaski, Ryan and Axelrod, Simon and Samsi, Siddharth and Gomez-Bombarelli, Rafael and Coley, Connor and Gadepally, Vijay},
    month = may,
    year = {2022}    
}

@article{huang_artificial_2022,
    title = {Artificial intelligence foundation for therapeutic science},
    volume = {18},
    copyright = {2022 Springer Nature America, Inc.},
    issn = {1552-4469},
    url = {https://www.nature.com/articles/s41589-022-01131-2},
    doi = {10.1038/s41589-022-01131-2},
    language = {en},
    number = {10},
    urldate = {2023-03-20},
    journal = {Nature Chemical Biology},
    author = {Huang, Kexin and Fu, Tianfan and Gao, Wenhao and Zhao, Yue and Roohani, Yusuf and Leskovec, Jure and Coley, Connor W. and Xiao, Cao and Sun, Jimeng and Zitnik, Marinka},
    month = oct,
    year = {2022},
    pages = {1033--1036},
}

@article{brown_guacamol_2019,
    title = {{GuacaMol}: {Benchmarking} {Models} for de {Novo} {Molecular} {Design}},
    volume = {59},
    issn = {1549-9596},
    shorttitle = {{GuacaMol}},
    url = {https://doi.org/10.1021/acs.jcim.8b00839},
    doi = {10.1021/acs.jcim.8b00839},
    number = {3},
    urldate = {2023-03-20},
    journal = {J. Chem. Inf. Model.},
    author = {Brown, Nathan and Fiscato, Marco and Segler, Marwin H.S. and Vaucher, Alain C.},
    month = mar,
    year = {2019},
    pages = {1096--1108},
}

@article{schwaller_mapping_2021,
    title = {Mapping the space of chemical reactions using attention-based neural networks},
    volume = {3},
    copyright = {2021 The Author(s), under exclusive licence to Springer Nature Limited},
    issn = {2522-5839},
    url = {https://www.nature.com/articles/s42256-020-00284-w},
    doi = {10.1038/s42256-020-00284-w},
    language = {en},
    number = {2},
    urldate = {2023-03-20},
    journal = {Nature Machine Intelligence},
    author = {Schwaller, Philippe and Probst, Daniel and Vaucher, Alain C. and Nair, Vishnu H. and Kreutter, David and Laino, Teodoro and Reymond, Jean-Louis},
    month = feb,
    year = {2021},
    pages = {144--152},
}

@article{zhu_torchdrug_2022,
    title = {{TorchDrug}: {A} {Powerful} and {Flexible} {Machine} {Learning} {Platform} for {Drug} {Discovery}},
    shorttitle = {{TorchDrug}},
    url = {http://arxiv.org/abs/2202.08320},
    urldate = {2023-03-20},
    journal = {arXiv:2202.08320 [cs.LG]},
    author = {Zhu, Zhaocheng and Shi, Chence and Zhang, Zuobai and Liu, Shengchao and Xu, Minghao and Yuan, Xinyu and Zhang, Yangtian and Chen, Junkun and Cai, Huiyu and Lu, Jiarui and Ma, Chang and Liu, Runcheng and Xhonneux, Louis-Pascal and Qu, Meng and Tang, Jian},
    month = feb,
    year = {2022}    
}

@misc{hu_strategies_2020,
    title = {Strategies for {Pre}-training {Graph} {Neural} {Networks}},
    url = {http://arxiv.org/abs/1905.12265},
    urldate = {2023-03-20},
    publisher = {arXiv},
    author = {Hu, Weihua and Liu, Bowen and Gomes, Joseph and Zitnik, Marinka and Liang, Percy and Pande, Vijay and Leskovec, Jure},
    month = feb,
    year = {2020}
}

@article{chithrananda_chemberta_2020,
    title = {{ChemBERTa}: {Large}-{Scale} {Self}-{Supervised} {Pretraining} for {Molecular} {Property} {Prediction}},
    shorttitle = {{ChemBERTa}},
    url = {http://arxiv.org/abs/2010.09885},
    doi = {10.48550/arXiv.2010.09885},
    urldate = {2023-03-20},
    publisher = {arXiv},
    journal = {arXiv:2010.09885 [cs.LG]},
    author = {Chithrananda, Seyone and Grand, Gabriel and Ramsundar, Bharath},
    month = oct,
    year = {2020}
}

@article{pedregosa_scikit-learn_2011,
    title = {Scikit-learn: {Machine} {Learning} in {Python}},
    volume = {12},
    issn = {1533-7928},
    shorttitle = {Scikit-learn},
    url = {http://jmlr.org/papers/v12/pedregosa11a.html},
    number = {85},
    urldate = {2023-03-20},
    journal = {Journal of Machine Learning Research},
    author = {Pedregosa, Fabian and Varoquaux, Gaël and Gramfort, Alexandre and Michel, Vincent and Thirion, Bertrand and Grisel, Olivier and Blondel, Mathieu and Prettenhofer, Peter and Weiss, Ron and Dubourg, Vincent and Vanderplas, Jake and Passos, Alexandre and Cournapeau, David and Brucher, Matthieu and Perrot, Matthieu and Duchesnay, Édouard},
    year = {2011},
    pages = {2825--2830},
}

\clearpage
\renewcommand{\thepage}{S\arabic{page}}
\renewcommand{\thesection}{S\arabic{section}}
\renewcommand{\thetable}{S\arabic{table}}
\renewcommand{\thefigure}{S\arabic{figure}}

\setcounter{page}{1}
\setcounter{figure}{0}
\setcounter{table}{0}

\newrefsection

{
\centering
\Huge Supporting Information \\ \bigskip
\LARGE \thetitle \\ \bigskip
\large
David E. Graff\textsuperscript{$\dagger, \ddagger$},
Edward O. Pyzer-Knapp\textsuperscript{\P},
Kirk E. Jordan\textsuperscript{\#},
Eugene I. Shakhnovich\textsuperscript{$\dagger$},
and Connor W. Coley\textsuperscript{$\ast, \ddagger$, \S} \\ \bigskip

\normalsize\itshape
\textsuperscript{$\dagger$}\ccb \\
\textsuperscript{$\ddagger$}\cheme \\
\textsuperscript{\upshape\P}\ibm \\
\textsuperscript{\#}\ibmtwo \\
\textsuperscript{\upshape\S}\eecs \\ \bigskip
\upshape \sffamily E-mail: ccoley@mit.edu \\
}

\section*{Additional Methods}

\begin{table}[h]
\centering
\begin{threeparttable}
\caption{Dimensionality of all representations tested}
\label{tbl:repr-sizes}
\begin{tabular}{@{}lr@{}}
    \toprule
    representation & dimensionality \\ \midrule
    descriptor & 14 \\
    FP & 512 \\
    VAE & 128 \\
    GIN & 300 \\
    ChemBERTa & 384 \\
    ChemGPT & 2048 \\
    random & 128 \\ \bottomrule
\end{tabular}
\end{threeparttable}
\end{table}

\section*{Additional Results}

\begin{figure}[h]
    \centering
    \includegraphics[width=\textwidth]{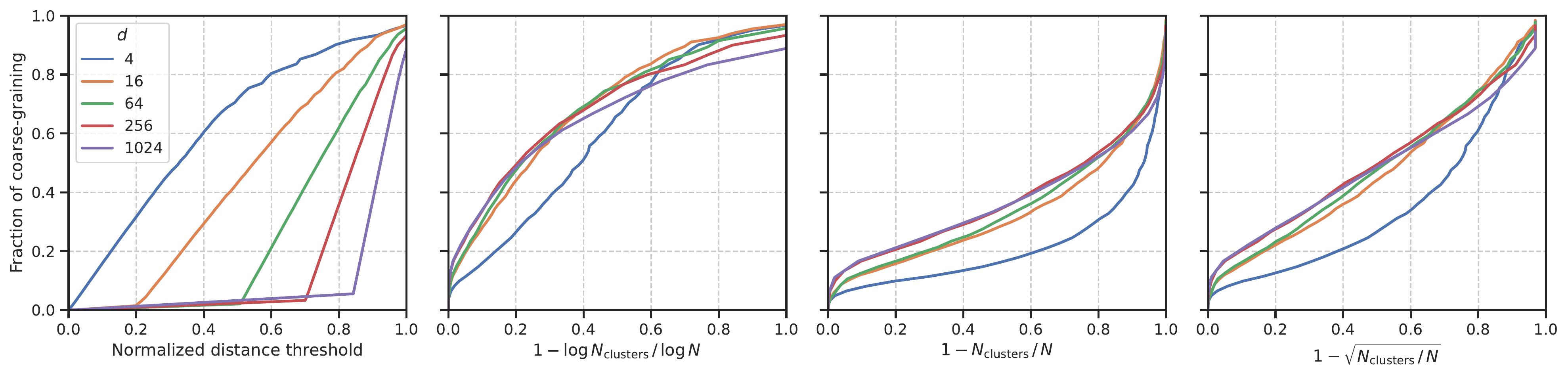}
    \caption{Examples of possible different formulations of the integration variable in the \newrogi{} formulation for \num{1000} points sampled from the domain $[0, 1]^d$. ``Fraction of coarse-graining'' is the fractional number of steps in the stepwise dendrogram of the clustering routine after subsampling. In all formulations aside from normalized distance threshold $t$, the curves of all datasets approximately overlap, indicating a domain of integration that is independent of representation dimensionality. However, only $1 - \log N_{\mathrm{clusters}} / \log N$ is defined over the constant domain $[0, 1]$. Note that the approximate overlap is caused by the dendrogram subsampling, but this step is performed for computational efficiency.}
    \label{fig:reformulation-ex}
\end{figure}

\begin{figure}
    \centering
    \includegraphics[width=0.9\textwidth]{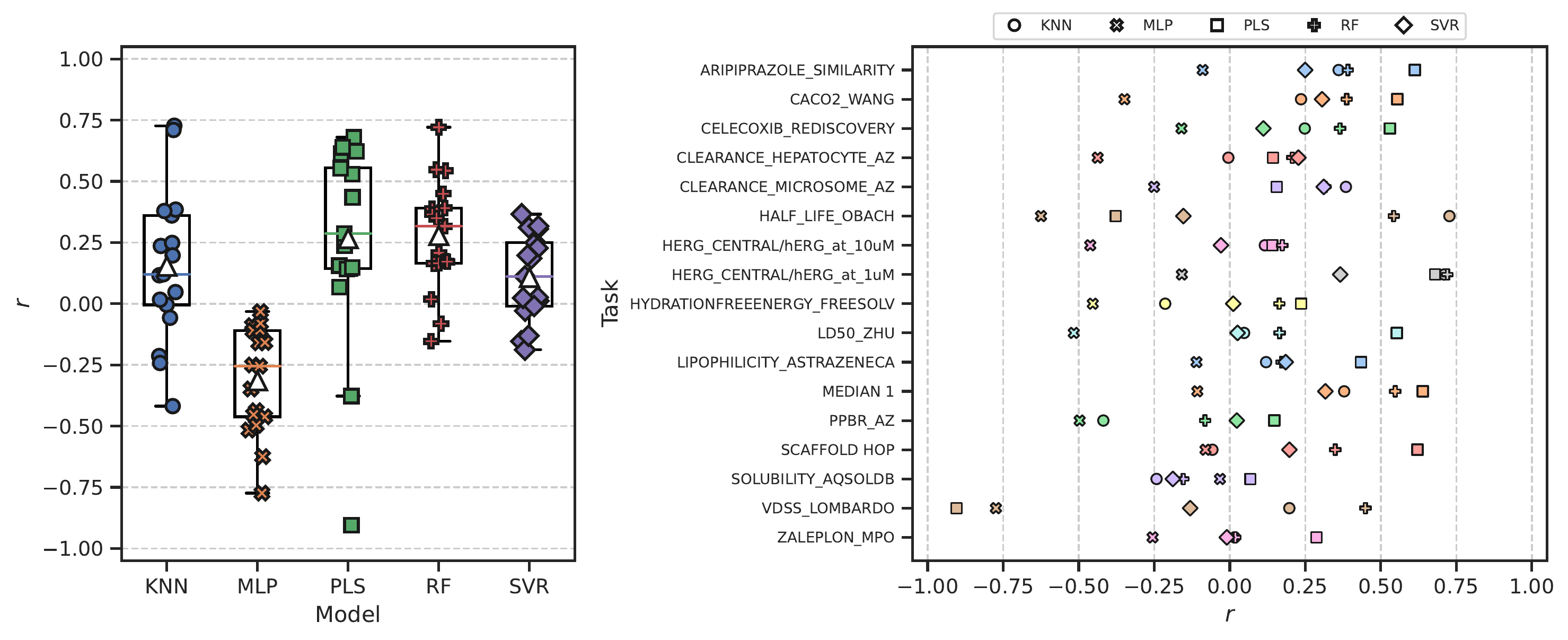}
    \caption{
        Pearson correlation coefficient $r$ between ROGI and cross-validated RMSE across all representations evaluated for a given pair of ML model and dataset.
        \textit{Left}: Box plot of correlations grouped by ML model architecture with individual data points plotted above. The median is depicted via the solid, colored line, and the mean by the white triangle ($\triangle$).
        \textit{Right}: Correlations grouped by dataset.
        \modelCaption.
    }
    \label{fig:model-task-v1}
\end{figure}

\begin{figure}
    \centering
    \includegraphics[width=0.9\textwidth]{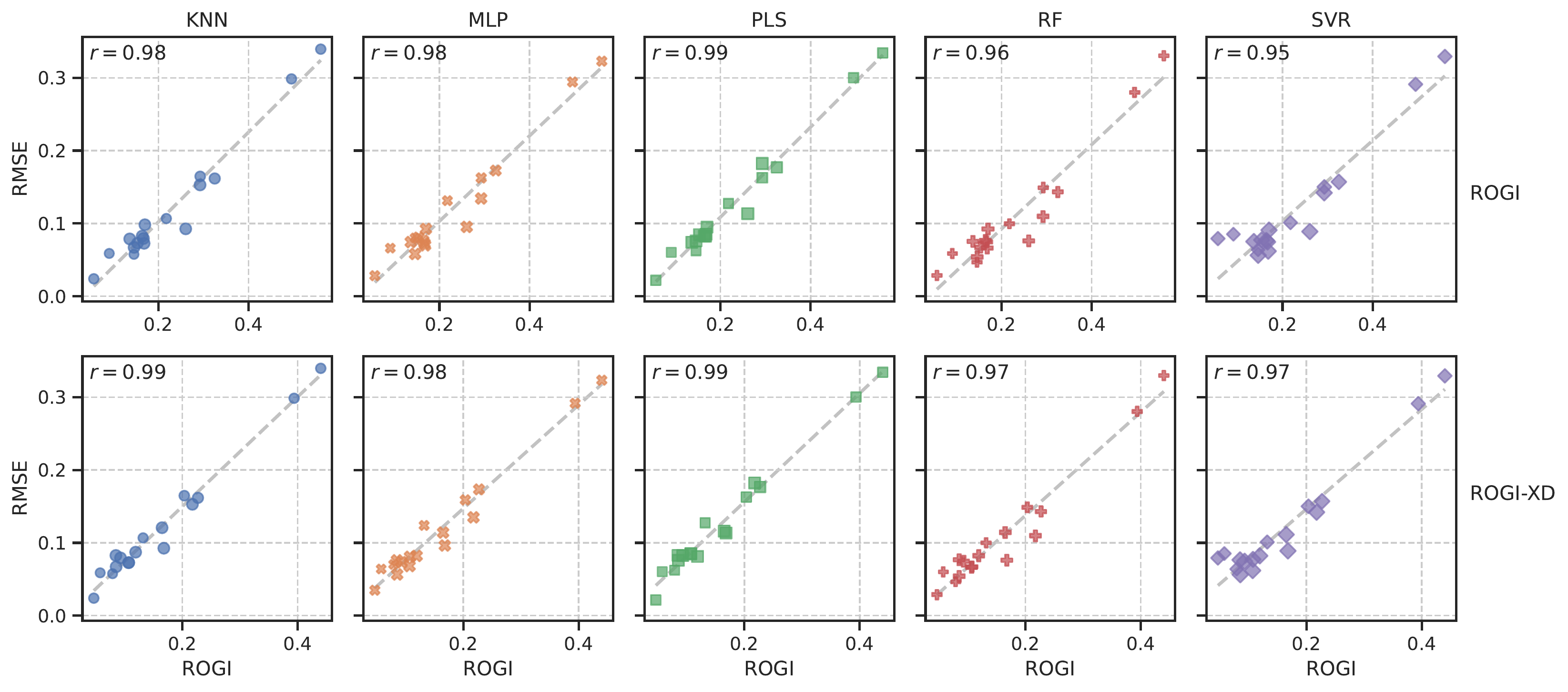}
    \caption{Roughness metric vs RMSE using descriptors across all datasets. Marker size is proportional to the natural logarithm of the dataset size. \modelCaption.}
    \label{fig:tasks-both}
\end{figure}

\begin{figure}
    \centering
    \includegraphics[height=0.925\textheight]{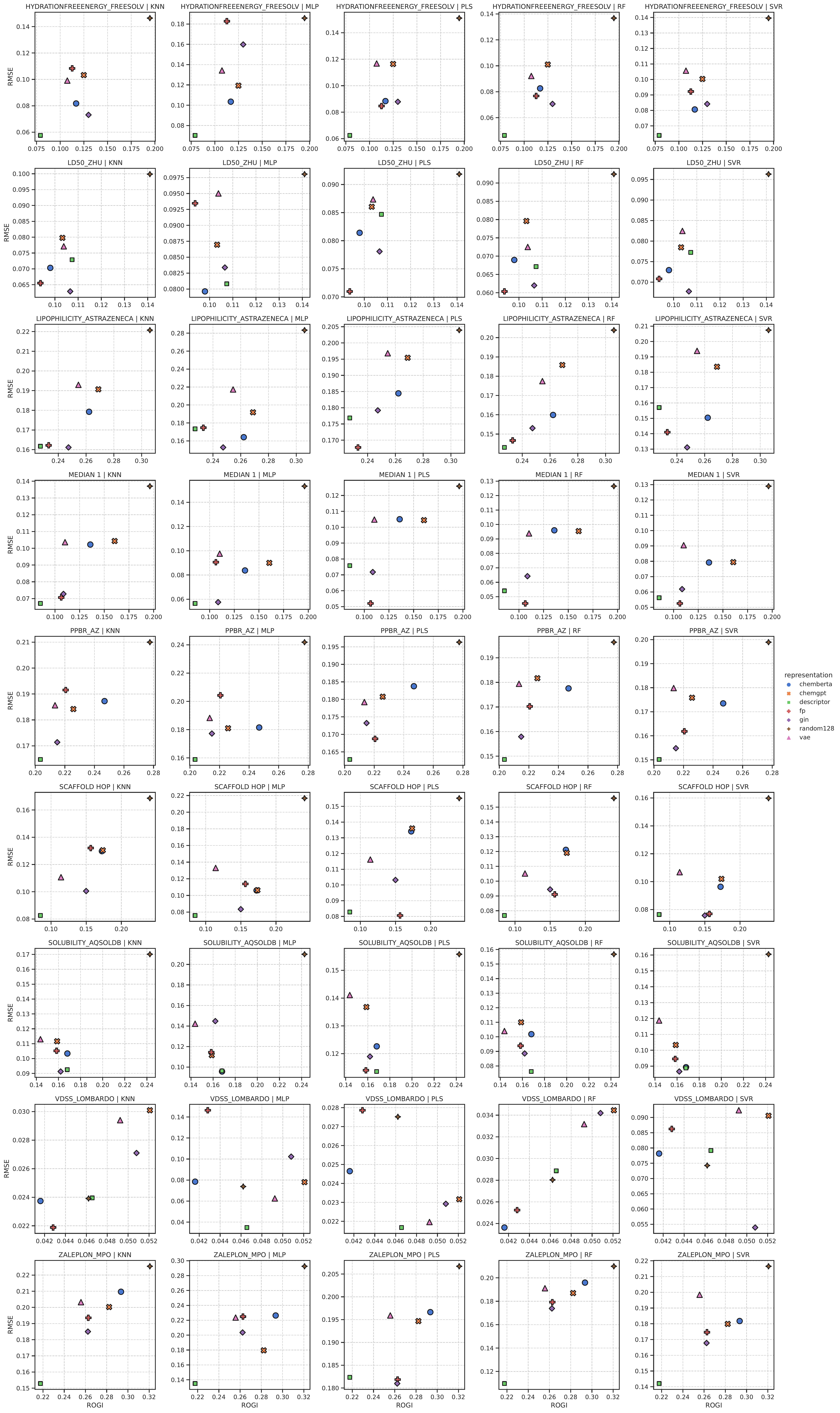}
    \caption{
        \newrogi{} vs. RMSE for each combination of task and ML model (1/2).
        \modelCaption; \reprCaption.
    }
    \label{fig:all-v2-1}
\end{figure}

\begin{figure}
    \centering
    \includegraphics[height=0.925\textheight]{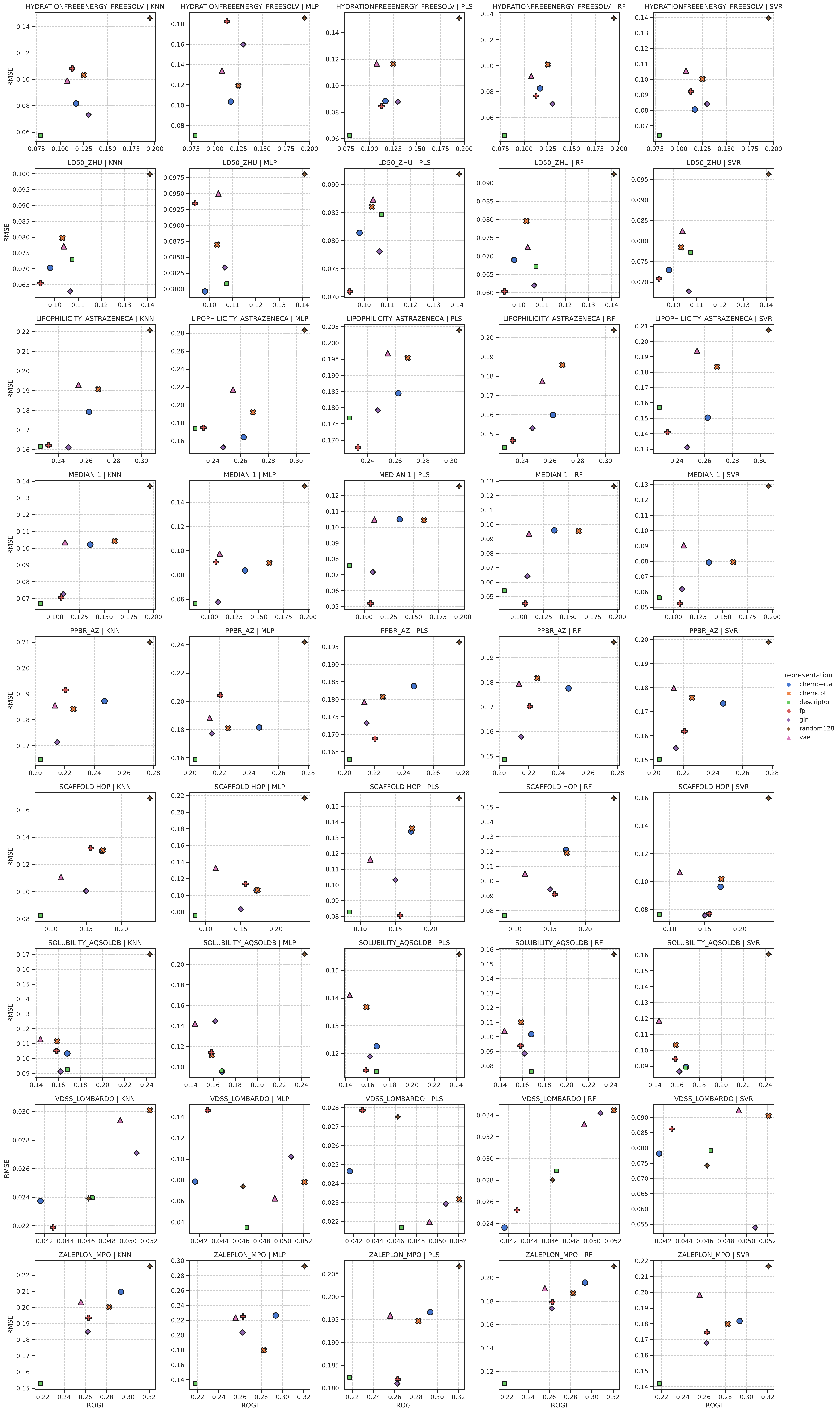}
    \caption{
        \newrogi{} vs. RMSE for each combination of task and ML model (2/2).
        \modelCaption; \reprCaption.
    }
    \label{fig:all-v2-2}
\end{figure}

\begin{figure}
    \centering
    \includegraphics[height=0.925\textheight]{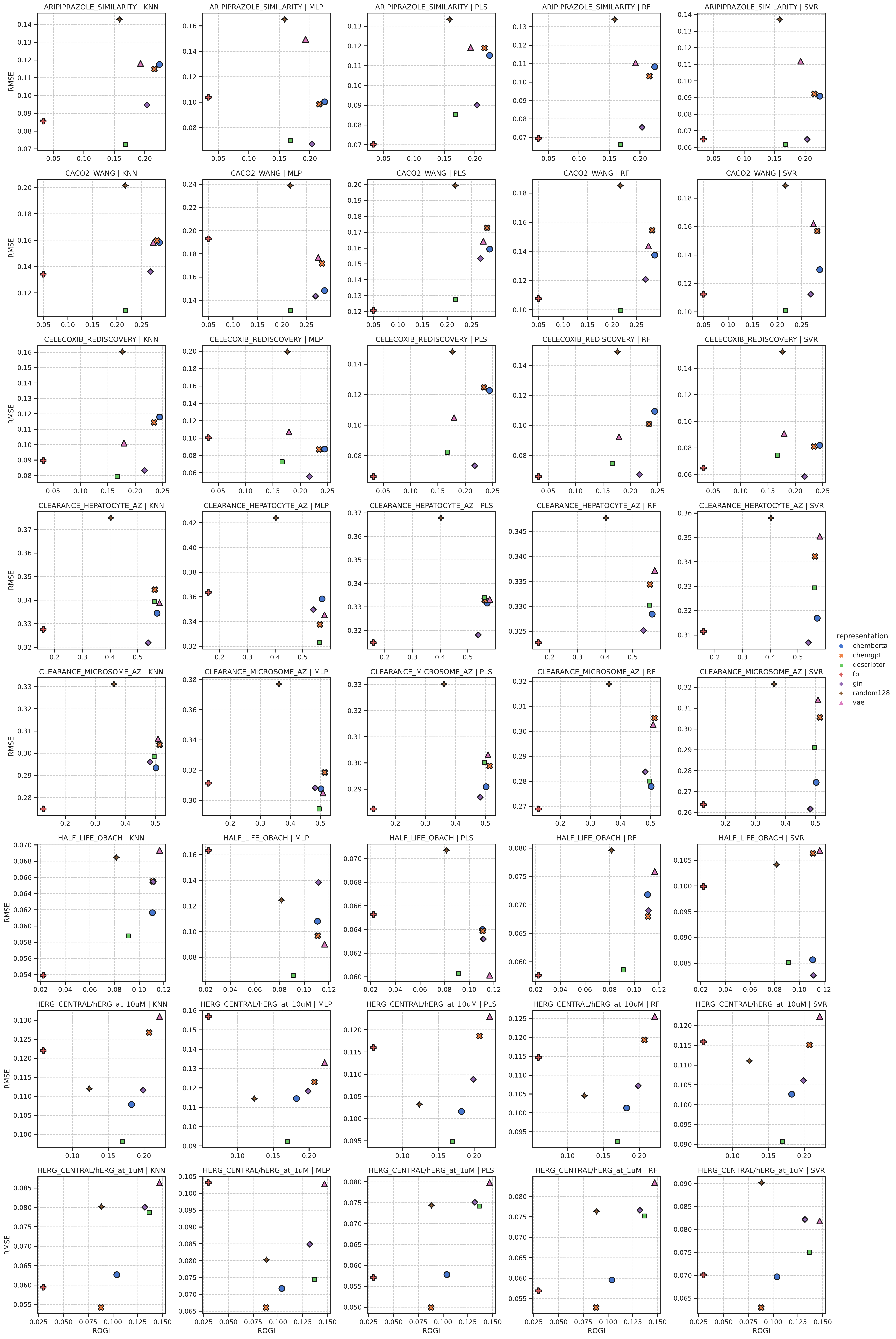}
    \caption{
        ROGI vs. RMSE for each combination of task and ML model (1/2).
        \modelCaption; \reprCaption.
    }
    \label{fig:all-v1-1}
\end{figure}

\begin{figure}
    \centering
    \includegraphics[height=0.925\textheight]{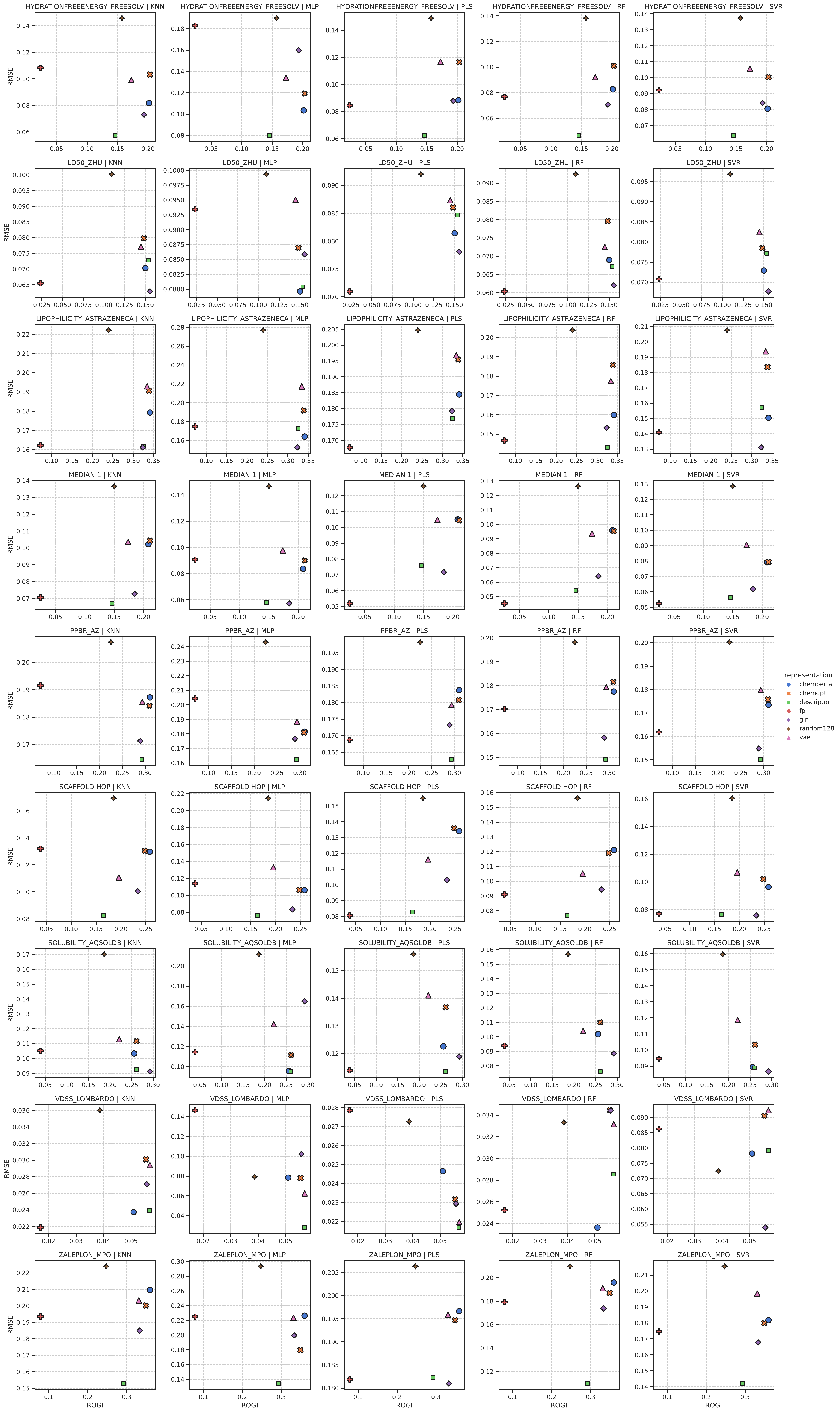}
    \caption{
        ROGI vs. RMSE for each combination of task and ML model (2/2).
        \modelCaption; \reprCaption.
    }
    \label{fig:all-v1-2}
\end{figure}

\begin{figure}[b!]
    \centering
    \includegraphics[width=0.6\textwidth]{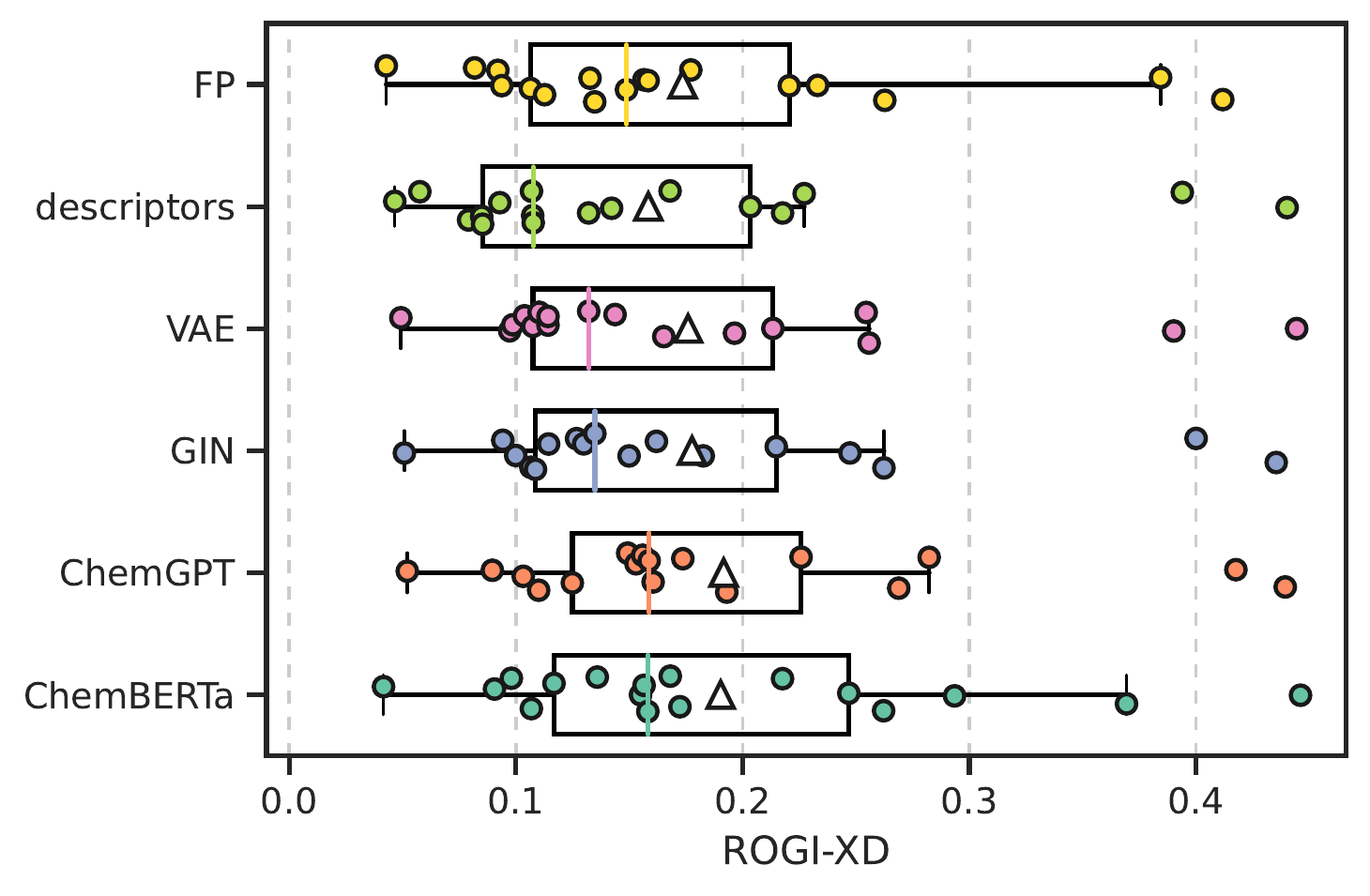}
    \caption{
        Distribution of \newrogi{} values on all tasks for the respective representation. The median is depicted via the solid, colored line, and the mean by the white triangle ($\triangle$).
        \reprCaption.
    }
    \label{fig:rogi-dist}
\end{figure}


\end{document}